\documentclass[twocolumn]{aastex701}

\usepackage{bm}
\usepackage{xspace}
\usepackage{xcolor}
\newcommand{\ms}{$\, \rm m \, s^{-1}$\xspace}

\shorttitle{MPM for Low-Velocity Asteroid Collisions}
\shortauthors{Fukuta et al.}
\submitjournal{ApJ Letters}
\usepackage{amsmath}
\usepackage{graphicx}
\usepackage{physics}

\begin{document}

\title{Application and Evaluation of the Material Point Method for Low-Velocity Asteroid Collisions Involving Large Deformations}

\author[orcid=0009-0007-9715-5162,gname=Haru,sname=Fukuta]{Haru Fukuta}
\affiliation{Department of Mathematical Information Science, Graduate School of Science and Engineering, Chiba University, 1-33 Yayoi-cho, Inage-ku, Chiba 263-8522, Japan}
\email{25wm0212@student.gs.chiba-u.jp}

\author[orcid=0000-0002-5316-9171,gname=Tomoaki,sname=Ishiyama]{Tomoaki Ishiyama} 
\affiliation{Information Strategy Organization, Chiba University, 1-33 Yayoi-cho, Inage-ku, Chiba 263-8522, Japan}
\email{ishiyama@chiba-u.jp}

\author[orcid=0000-0003-4366-6518,gname=Shu-ichiro,sname=Inutsuka]{Shu-ichiro Inutsuka}
\affiliation{Department of Physics, Nagoya University, Furo-cho, Chikusa-ku, Nagoya, Aichi 464-8602, Japan}
\email{inutsuka@nagoya-u.jp}



\begin{abstract}

The collisional lifetimes of asteroids larger than $\sim 10$ km in the present-day main belt are estimated to exceed the age of the solar system, suggesting that their shapes may preserve memories of the collisional environment in the primordial solar system. Recent studies performed systematic simulations of equal-mass and low-velocity ($50$--$400\,\mathrm{ms}^{-1}$) collisions between 50-km-radius asteroids using Smoothed Particle Hydrodynamics (SPH), identifying the impact conditions that produce irregular shapes. 
As an alternative approach, we here employ the Material Point Method (MPM), a particle-grid method well suited to large deformation of solids and free from tensile instability, and we assess its applicability to such systems by direct comparison with the established SPH results.
We integrate a previously established rock model combined with a fracture model, the Tillotson equation of state, and Drucker-Prager friction, with the efficient non-associative Drucker-Prager plasticity solver originally developed for granular flows.
We perform 135 collision simulations of equal-mass rocky asteroids spanning 15 impact velocities ($50$--$400\,\mathrm{ms}^{-1}$) and 9 impact angles ($5$--$45^\circ$), following the setup of the previous SPH simulations. The mass, axis ratio, and shape classification of the largest remnants closely reproduce the trends of the previous SPH simulations. Since most material transitions into a granular state early in the collision, this agreement appears to depend primarily on the validity of the frictional treatment of the granular material, rather than on the details of the fracture model. These results demonstrate the applicability of MPM to low-velocity asteroid collision simulations and support cross-validation between MPM- and SPH-based approaches.

\end{abstract}


\section{Introduction}

In the current main belt, the frequency of catastrophic collisions between celestial bodies larger than $10$~km in diameter is extremely low, with collision lifetimes estimated to exceed $10$ billion years~\citep{O'Brien2005}. Hence, the irregular shapes of asteroids observed today are thought to preserve memories of the violent collisional environment of the early solar system, potentially reflecting the specific impact conditions during their formation process.

Based on this background,~\citet{Sugiura2018} conducted systematic simulations of equal-mass asteroid impacts with radii of $50\,\mathrm{km}$ under low impact velocities ($50$--$400\,\mathrm{ms}^{-1}$) and a range of impact angles ($5$--$45^\circ$). They employed the Smoothed Particle Hydrodynamics (SPH) method and identified the range of initial conditions under which the resulting remnants acquired irregular shapes. However, most previous studies have relied on SPH or $N$-body methods~\citep{Michel2013, Jutzi2015b, Jutzi2017, Schwartz2018, Sugiura2018, Sugiura2019, Sugiura2020, Kurosaki2026}. As Lagrangian approaches, these methods can suffer from increasing computational costs when particle separations become small. In addition, SPH is known to exhibit tensile instability. 
Independent verification using a methodologically distinct algorithm therefore provides a valuable consistency check, irrespective of the specific limitations of either approach.

A promising alternative is the Material Point Method \citep[MPM;][]{Sulsky1995}, a particle-based numerical technique developed as an extension of the Particle-In-Cell (PIC) method and designed primarily for handling large deformations in solid bodies. In this method, material points carry mass and momentum, whereas spatial derivatives and momentum exchange are computed on  a fixed background grid. Modern MPM implementations adopt the deformation-gradient-based constitutive formulations~\citep{Mast2013, Mast2014}, in which constitutive response is consistently derived from a strain energy density function of the deformation gradient. Combined with objective stress-updating used in nonlinear continuum mechanics, this formulation maintains consistency between strain and energy even in long-term simulations involving large deformations. Separately, a direct comparison between MPM and SPH in hypervelocity impact problems has shown that MPM offers several advantages over SPH~\citep{Ma2009}. Especially, MPM does not suffer from tensile instability that corresponds to unphysical particle clumping in a negative pressure region generically found in the standard SPH method~\citep[see also][for the resolution in Godunov SPH]{Sugiura2016}.

Modern MPM implementations have advanced progressively in recent years by developments of higher-order particle-grid spatial interpolation, including B-spline-based approaches~\citep{Michael2008}, and improved particle-grid transfer schemes~\citep{Jiang2015, Fu2017}, improving numerical stability for simulations with large deformations.
From an algorithmic perspective, MPM combines advantages of both mesh-based and particle-based methods, in which the background grid is reset every timestep, thereby avoiding mesh distortion under large deformations as originally emphasized by~\citet{Sulsky1995}. 
In contrast to SPH, MPM evaluates physical quantities through a background grid, which does not require neighbor-particle searching during contact, and also eliminates free-surface tracking. 
Despite these advantages, applications of MPM to asteroid-scale impact problems remain comparatively limited, motivating a direct benchmark against established SPH results.

Of particular importance in the context of asteroid impacts is the treatment of granular materials;~\citet{Klar2016} established highly efficient techniques for handling granular media within the MPM framework based on the Drucker-Prager model~\citep[DP Model; ][]{Drucker1952}. By combining with the ASFLIP time integrator \citep{Fei2021}, which is drived from the FLIP~\citep{Zhu2005} and APIC~\citep{Jiang2015} methods, the numerical dissipation inherent in PIC is reduced by FLIP, while the numerical noise of FLIP is further reduced by APIC, highlighting the potential of MPM for asteroid simulations. 

Building on these developments, MPM has also begun to be applied to planetary science problems in recent years. \citet{ElMir2019} developed a hybrid MPM--$N$-body framework for hypervelocity asteroid impacts and subsequent gravitational reaccumulation, while \citet{Yan2026} recently applied MPM to hypervelocity impacts on asteroids. Nevertheless, applications of MPM to asteroid impacts remain in their early stages, and studies of large deformations in equal-mass asteroid collisions are still limited.

In this paper, we perform MPM simulations similar to the SPH simulations of equal-mass asteroid impacts with low impact velocities performed by~\citet{Sugiura2018}, 
and compare the MPM results with their SPH results to assess their consistency.
This paper is organized as follows: Section~\ref{sec:method} describes the numerical methods, Section~\ref{sec:setup} presents the simulation setup, Section~\ref{sec:results} presents the results, and Section~\ref{sec:discussion} provides discussion and conclusions.


\section{Overview of Material Point Method}\label{sec:method}
The computational cycle of the MPM employed in this study is outlined below. Here, the subscripts $\cdot_p$ and $\cdot_i$ denote the physical quantities associated with particles and grids, respectively, while the superscript $\cdot^n$ indicates the time step. A hat ($\hat{\cdot}$) and a tilde ($\tilde{\cdot}$) denote trial quantities prior to plasticity correction and quantities interpolated from the grid to the particles, respectively.
Since a leapfrog scheme is used for the time integration, particle and grid velocities, and the velocity gradient are evaluated at half time steps ($n+\frac12$), whereas all other quantities are evaluated at integer time steps.
The correspondence between the present MPM/ASFLIP formulation and the standard SPH governing equations for elastic bodies is summarized in Appendix~\ref{sec:CS}.

\begin{enumerate}
\item Stress calculation per particle: A trial Kirchhoff stress tensor
($\hat{\mathit{\tau}}^{n}_p$) is computed from the deviatoric stress derived
from the trial elastic deformation gradient tensor
($\hat{\mathit{F}}^{\mathrm{e}}_p$) and the pressure obtained from
the Tillotson equation of state (EOS), proposed by~\citet{Tillotson1962},
together with an artificial viscosity contribution (see Appendix~\ref{sec:PS}
and Appendix~\ref{sec:visc}).

\item Apply plasticity: Based on the trial elastic deformation gradient
tensor ($\hat{\mathit{F}}^{\mathrm{e}}_p$), the specific internal
energy ($E_{p}^{n-1}$), and the damage variable ($D_p^n$), the
rock constitutive model is applied to the trial stress
($\hat{\mathit{\tau}}^{n}_p$) to obtain the corrected elastic deformation
gradient tensor ($\mathit{F}^{\mathrm{e},n}_p$) and the corrected Kirchhoff
stress tensor ($\mathit{\tau}^{n}_p$), using the interpolation scheme of
\citet{Lundborg1968, Collins2004, Jutzi2015} to smoothly transition the
yield surface from intact rock to granular material (see Appendix
\ref{sec:AP}). The specific internal energy ($E_{p}^{n}$) is
then updated from the corrected Kirchhoff stress tensor ($\mathit{\tau}^{n}_p$).

\item Particles to grids (P2G): Particle masses ($m_p$), velocities ($\bm{v}^{n-\frac{1}{2}}_p$), velocity gradients ($\nabla \bm{v}^{n-\frac{1}{2}}_p$), and Kirchhoff stresses ($\mathit{\tau}^{n}_p$) are interpolated to background grid masses ($m^n_i$), momenta ($(m\bm{v})^{n-\frac{1}{2}}_i$), and forces ($\bm{f}^{n}_i$). The grid forces include both the internal force derived from the particle stresses and external body forces such as gravity.

\item Grid update: Grid velocities ($\bm{v}^{n+\frac{1}{2}}_i$) are updated using grid forces ($\bm{f}^{n}_i$) and a lumped mass matrix.

\item Grids to particles (G2P): Pre-update (interpolated) grid velocities ($\bm{v}^{n-\frac{1}{2}}_i$) and updated grid velocities ($\bm{v}^{n+\frac{1}{2}}_i$) are interpolated back to the particles at their positions ($\bm{x}^{n}_p$) to obtain the interpolated particle velocities ($\tilde{\bm{v}}^{n-\frac{1}{2}}_p$ and $\tilde{\bm{v}}^{n+\frac{1}{2}}_p$) and velocity gradients ($\nabla\tilde{\bm{v}}^{n+\frac{1}{2}}_p$).

\item Particle update: Using the interpolated particle velocity $\tilde{\bm{v}}^{n-\frac{1}{2}}_p$, the interpolated particle velocity $\tilde{\bm{v}}^{n+\frac{1}{2}}_p$ and the particle position $\bm{x}^n_p$ are corrected via the APIC-Style FLIP scheme~\citep[ASFLIP;][]{Fei2021}, where the ASFLIP parameters are set to $\alpha=0.995$ and $\beta=1.0$, yielding the updated particle velocity ($\bm{v}^{n+\frac{1}{2}}_p$) and position ($\bm{x}^{n+1}_p$), respectively. The trial elastic deformation gradient tensor ($\hat{\mathit{F}}^{\mathrm{e}}_p$) is obtained from the velocity gradient ($\nabla\tilde{\bm{v}}^{n+\frac{1}{2}}_p$).

\item Damage update: The evolution of fracture in each particle is described by
a damage variable $D \in [0,1]$, based on the model proposed by
\citet{Grady1980}, as implemented and extended for SPH by
\citet{Benz1995, Benz1999}. The damage variable ($D_p^{n+1}$) is updated based on
the maximum tensile principal Hencky strain computed from the total
deformation gradient tensor
($\mathit{F}^{n}_p = \hat{\mathit{F}}^{\mathrm{e}}_p\mathit{F}^{\mathrm{p},n-1}_p$, with $\mathit{F}^{\mathrm{p},n-1}_p$ denoting the plastic
deformation gradient obtained in the preceding time step; see Appendix
\ref{sec:AP}).

\end{enumerate}
Further details of these standard MPM processes can be found in the literature~\citep[e.g.,][]{Jiang2015, Hu2018, Klar2016, Fei2021}.

\section{Numerical Setup}\label{sec:setup}

The numerical setup is based on SPH simulations of equal-mass asteroid impacts with low impact velocities conducted in~\citet{Sugiura2018}, which enables us to compare SPH and MPM qualitatively. 
Each asteroid is modeled as uncompressed basalt spheres with uniform density $\rho = 2700\,\mathrm{kg}\,\mathrm{m}^{-3}$ and a radius of $R = 50\,\mathrm{km}$, corresponding to the initial mass of $M_{\mathrm{target}}\approx 1.414\times 10^{18}\,\mathrm{kg}$.

The total number of particles was set to $94,648$ ($47,324$ particles per asteroid). The background grid cell width was set to $dx = 2.5\,\mathrm{km}$.
The initial particle distribution was approximately uniform, which is sufficient for MPM since the P2G interpolation minimizes sensitivity to exact uniformity. 

The parameters $v_{\mathrm{imp}}$ and $\theta_{\mathrm{imp}}$ are defined as the impact velocity and angle relative to the line of center at the moment of closest approach (center-to-center distance $= 2R$). The parameter space consists of a $15\times9$ grid covering velocities from 50 to 400\ms in 25\ms increments and angles from 5 to $45^\circ$ in 5$^\circ$ increments, yielding 135 cases. The initial separation of two asteroids was $4R$, and the initial equal-magnitude opposite velocity vectors were back-calculated using two-body approximation from the impact condition. Following the previous study, the total simulation time was set to $T = 1.0 \times 10^5\,\mathrm{s}$, corresponding to 819,200 time steps ($dt\simeq0.122\,\mathrm{s}$), to ensure collisional reaccumulation is sufficiently completed. 

To reduce computational costs, gravitational acceleration was updated once every 32 MPM steps using a second-order Runge--Kutta method. The computed acceleration was held constant until the next gravitational update and transferred to the background grid along with the internal forces during the P2G transfer. Particle interactions were treated as point-mass interactions with a gravitational softening length of $0.05dx=0.125\,\mathrm{km}$.

To directly compare previous and current studies, we adopt the same analysis method for the shape of largest remnant
used in ~\citet{Sugiura2018}. The largest remnants were identified using a friends-of-friends algorithm~\citep{Huchra1982} with the threshold distance of 3 cell widths. We quantified the remnant morphology using the axis ratios of $b/a$ and $c/a$, 
where $a, b$, and $c$ denote the major, intermediate, and minor axial lengths, respectively. We denote the remnant mass as $M_{\mathrm{lr}}$. 
All simulations were performed on the GPU cluster at the Center for Computational Astrophysics (CfCA), National Astronomical Observatory of Japan, using a single NVIDIA A100 GPU. The MPM solver used in this study is optimized on GPU systems, employing a GPU-oriented memory layout inspired by the memory-efficient scheme of \citet{Gao2018b}.

\section{Numerical Result}\label{sec:results}
\begin{figure*}[ht!]
  \plotone{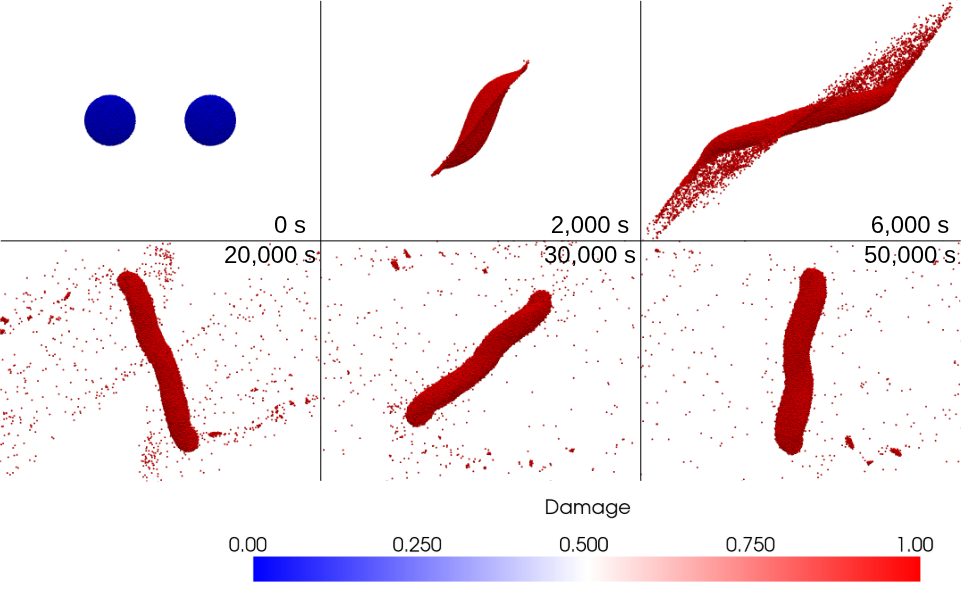}
  \caption{Snapshots of an oblique collision simulation ($v_{\mathrm{imp}} = 200$\ms and $\theta_{\mathrm{imp}} = 15^\circ$). Following the impact, part of the material is ejected from the contact region, while the material near the impact point remains coherent and begins to rotate, forming the largest remnant. A fraction of the ejecta trails behind the rotating remnant, following its major-axis tips and forming distinct tails that gradually disperse by approximately $2.0\times10^4 \, \mathrm{s}$. Thereafter, the remaining surrounding particles are slowly re-accreted onto the largest remnant until $5.0\times10^4 \, \mathrm{s}$.}
  \label{fig:irregular_shape}
\end{figure*}

Figure~\ref{fig:irregular_shape} represents snapshots of the collision simulation
with $v_{\mathrm{imp}} = 200$\ms and $\theta_{\mathrm{imp}} = 15^\circ$.
The largest remnant exhibits a highly elongated 
shape with $b/a = 0.23$, in good agreement with the value of $b/a 
\approx 0.2$ reported by~\citet{Sugiura2018}, confirming that MPM 
captures the same qualitative morphological outcomes as SPH under these 
impact conditions.

\begin{figure}[ht!]
  \centering
  \includegraphics[width=1.1\linewidth]{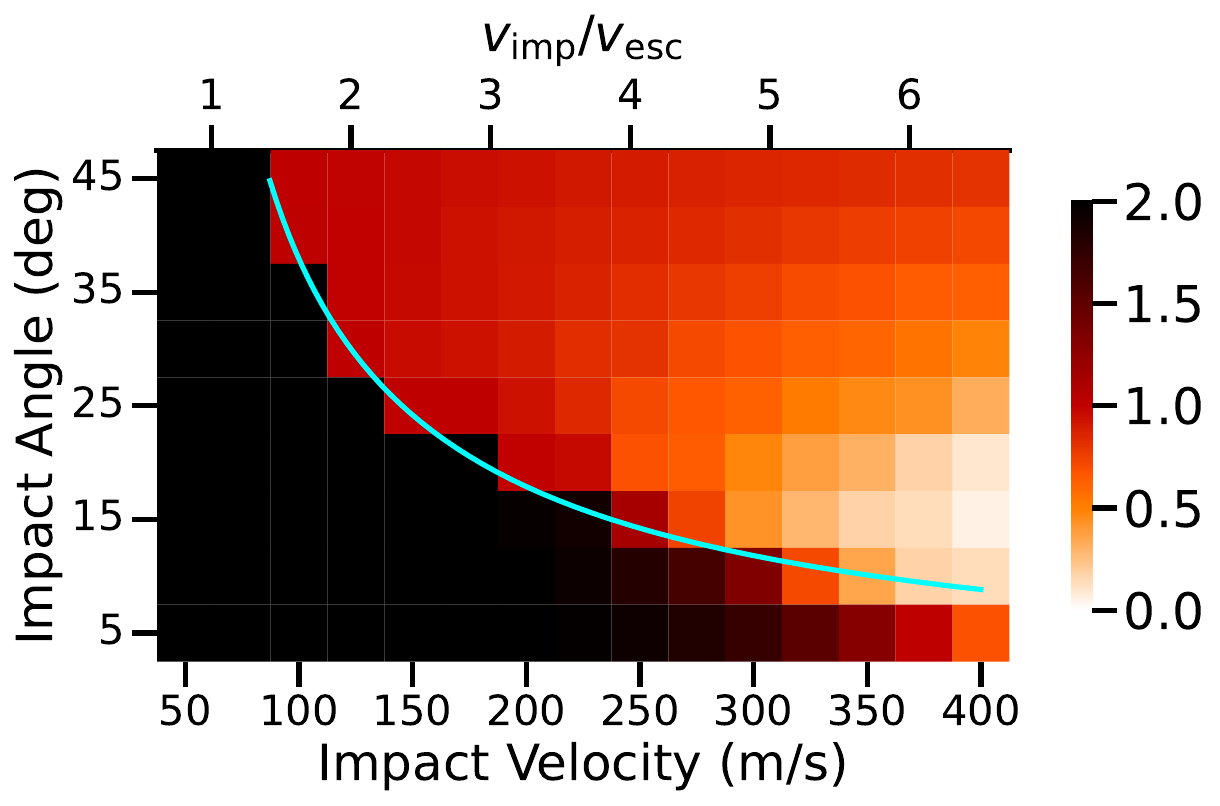} 
  \caption{Dependence of the normalized mass of the largest remnant, $M_{\mathrm{lr}}/M_{\mathrm{target}}$, on $v_{\mathrm{imp}}$ and $\theta_{\mathrm{imp}}$. Cyan curve indicates the condition $v_{\mathrm{imp}}\sin\theta_{\mathrm{imp}}=v_{\mathrm{esc}}$.}
  \label{fig:massheatmap}
\end{figure}

\begin{figure}[ht!]
  \centering
  \includegraphics[width=1.0\linewidth]{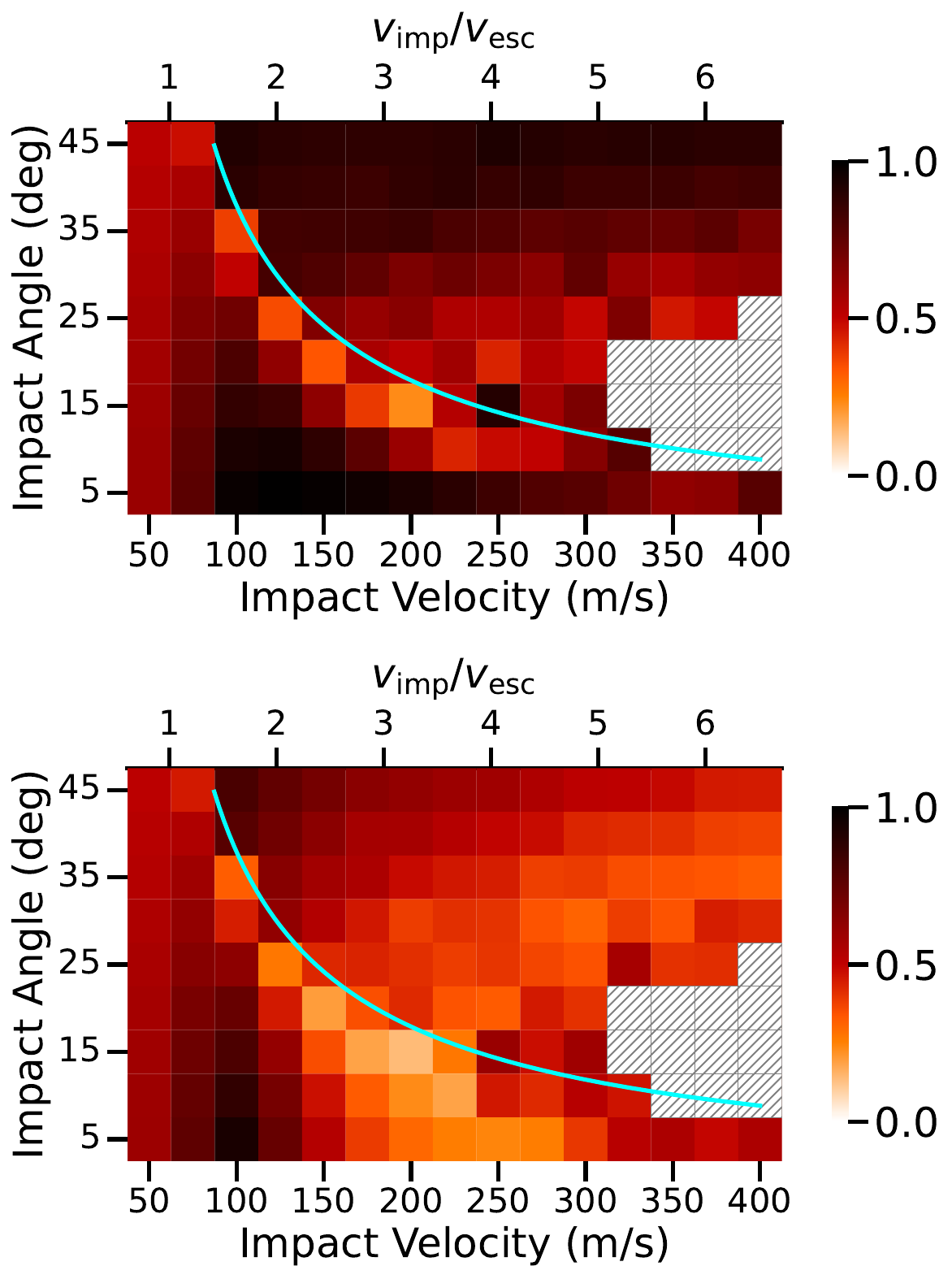} 
  \caption{Dependence of the axis ratios of the largest remnant, $b/a$ (top) and $c/a$ (bottom), on $v_{\mathrm{imp}}$ and $\theta_{\mathrm{imp}}$. Both ratios reach their minimum values near the regime boundary expressed by $v_{\mathrm{imp}} \sin\theta_{\mathrm{imp}} = v_{\mathrm{esc}}$ (cyan curves). The hatched region indicates cases where the largest remnant is classified as super‑catastrophic.}
  \label{fig:shapeheatmap}
\end{figure}

Figure~\ref{fig:massheatmap} shows the mass of the largest remnant ($M_{\mathrm{lr}}$) normalized by the initial mass of each asteroid ($M_{\mathrm{target}}$), as a function of impact velocity ($50 \le v_{\mathrm{imp}} \le 400$\ms) and impact angle ($5^\circ \le \theta_{\mathrm{imp}} \le 45^\circ$).
The cyan curve indicates the condition $v_{\mathrm{imp}}\sin\theta_{\mathrm{imp}} = v_{\mathrm{esc}}$, where $v_{\mathrm{esc}}$ is the mutual escape velocity of the two-body system. 
Generally, compared to the SPH method, MPM is prone to higher numerical dissipation because linear and angular momentum are conserved only in  a weak form rather than through pairwise interactions, which would reduce the post-impact tangential velocity. Consequently, it is anticipated that the transition boundary between merging and hit-and-run outcomes would shift toward the higher-velocity regime relative to the gravity-controlled threshold represented by the cyan curve. However, no significant shift from the cyan curve is observed in the present study, deviating only at the highest velocities ($\gtrsim 275$\ms), where catastrophic disruption occurs and the merging/hit-and-run distinction itself becomes meaningless. Overall, the resulting transition region is broadly consistent with the results of~\citet{Sugiura2018}.

Since this transition is primarily governed by the competition between the impact velocity and mutual gravitational attraction, this agreement indicates that the present implementation appropriately reproduces the gravity-dominated collision dynamics, including momentum transfer and gravitational binding of the colliding bodies. On the other hand, the largest-remnant mass alone does not provide sufficient evidence to assess the validity of the constitutive response, such as the Tillotson EOS or the DP plasticity model. Their influence is instead examined in the following sections through the axis ratio analyses and shape classification.

In low-angle (near-head-on) collisions, $M_{\mathrm{lr}}$ decreases 
gradually with increasing impact velocity, as the fraction of particles 
exceeding the escape velocity increases with the collision energy. In contrast, for high-angle (grazing) collisions, the two bodies merge into a single remnant below a critical impact velocity.
However, above this threshold, the bodies separate after only partial disruption, causing an abrupt transition between the merging and hit-and-run outcomes and a 
sharp drop in $M_{\mathrm{lr}}$. Furthermore, the significant decrease in $M_{\mathrm{lr}}$ ($M_{\mathrm{lr}} < 0.4M_{\mathrm{target}}$) due to 
catastrophic disruption is observed in high-velocity and low-angle collision regimes (300\ms $\le v_{\rm imp}$ and $\theta_{\rm imp} \le 25^\circ$), which is also consistent with the SPH simulations.

Figure~\ref{fig:shapeheatmap} shows the axis ratios (intermediate-to-major and minor-to-major) of the largest remnant as a function of impact velocity and angle. Both ratios reach their minimum values near the cyan curve ($v_{\mathrm{imp}} \sin\theta_{\mathrm{imp}} = v_{\mathrm{esc}}$), indicating that the most elongated shapes are generated in this boundary region. This suggests that the largest remnants undergo the strongest one-dimensional stretching immediately before the collision outcome transitions from merging to hit-and-run, as also seen in Figure~\ref{fig:massheatmap}.
For nearly head-on impacts ($\theta_{\mathrm{imp}} \approx 5^\circ$), the morphology exhibits a gradual evolution with increasing impact velocity. At $v_{\rm imp} \sim 250$\ms, $b/a \approx 1$ and $c/a$ attains its minimum, corresponding to flattened structures. In the higher-velocity regime ($\gtrsim 275$\ms), $c/a$ gradually increases while $b/a$ decreases as the impact velocity increases, indicating a progressive loss of planar flattening corresponding to elongated remnants that transition into super-catastrophic disruption.

\begin{figure}[ht!]
  \centering
  \includegraphics[width=1.1\linewidth]{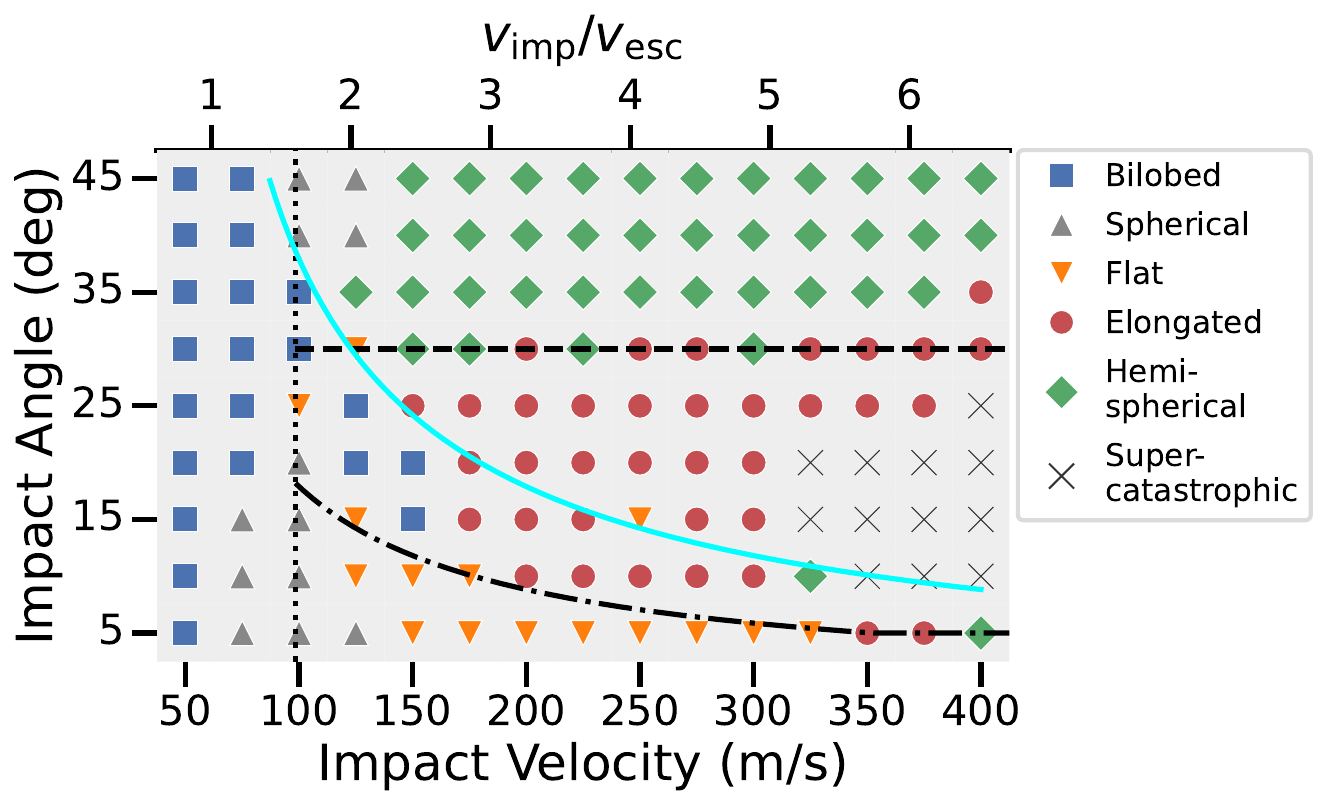} 
  \caption{Dependence of the resulting shape on $v_{\mathrm{imp}}$ and $\theta_{\mathrm{imp}}$. Dotted, dashed, and chain curves correspond to $v_{\mathrm{imp}}=1.6v_{\mathrm{esc}}$, $\theta_{\mathrm{imp}}=30^\circ$, and $v_{\mathrm{imp}}\sin\theta_{\mathrm{imp}}=0.5v_{\mathrm{esc}}$, respectively. Cyan curve indicates the condition $v_{\mathrm{imp}}\sin\theta_{\mathrm{imp}}=v_{\mathrm{esc}}$. 
  Each symbol shows the morphology: Bilobed ($b/a<0.7$,$c/a<0.7$,$M_{\mathrm{lr}}/M_{\mathrm{target}}=2.0$), Spherical ($b/a>0.7$,$c/a>0.7$), Flat ($b/a>0.7$,$c/a<0.7$,$M_{\mathrm{lr}}/M_{\mathrm{target}}>1.0$), 
  Elongated ($b/a<0.7$,$c/a<0.7$,$M_{\mathrm{lr}}/M_{\mathrm{target}}<2.0$), Hemispherical ($b/a>0.7$,$c/a<0.7$,$M_{\mathrm{lr}}/M_{\mathrm{target}}<1.0$), and 
  Super-catastrophic ($M_{\mathrm{lr}}/M_{\mathrm{target}}<0.4$). For detailed physical definitions, see~\citet{Sugiura2018}.}
  \label{fig:classheatmap}
\end{figure}

These results are summarized in Figure~\ref{fig:classheatmap}. The criteria for this shape classification are based on those defined in Table 1 of~\citet{Sugiura2018}.  For impacts with $v_{\mathrm{imp}} \le 150$\ms, bilobed largest remnants are generated in the high-angle regime, appearing at $\theta_{\mathrm{imp}} \gtrsim 20^\circ$ for $50$\ms $\le v_{\mathrm{imp}} \le 100$\ms and between the chain curve ($v_{\mathrm{imp}}\sin\theta_{\mathrm{imp}}=0.5v_{\mathrm{esc}}$) and the dashed line ($\theta_{\mathrm{imp}}=30^\circ$) for $100$\ms $\le v_{\mathrm{imp}} \le 150$\ms. Spherical ones are formed below the bilobed region at $75$\ms $\le v_{\mathrm{imp}} \le 100$\ms and sporadically at $\theta_{\mathrm{imp}} \gtrsim 40^\circ$ for $100$\ms $\le v_{\mathrm{imp}} \le 125$\ms. In the region between the chain curve and the dashed line, elongated shapes become dominant for $v_{\mathrm{imp}} \gtrsim 150$\ms, with the dominant morphology switching from bilobed to elongated around $150$\ms. Beyond the dotted line ($v_{\mathrm{imp}} \ge 1.6\,v_{\mathrm{esc}}$), the chain curve separates the regions where elongated or bilobed shapes and flat geometries are preferentially generated, with elongated or bilobed shapes appearing above the curve and flat shapes below it. In the region below this boundary, the flattened geometry can no longer be maintained at high velocities ($v_{\mathrm{imp}} \ge 350$\ms), consistent with the progressive loss of planar flattening noted above for $\theta_{\mathrm{imp}} \approx 5^\circ$ beyond $v_{\rm imp} \sim 250$\ms. On the other hand, the classification only changes visibly beyond $350$\ms because the classification itself is threshold-based. This is physically attributed to the conversion of impact energy into the rotational energy of the largest remnant during the re-accretion of dispersed material, where the increased contribution of re-accreted material progressively obscures the aggregation axis, preventing the stabilization of a clear aggregation axis.
Note that in the regime where $\theta_{\mathrm{imp}} \gtrsim 30^\circ$ and $v_{\mathrm{imp}} \gtrsim 125$\ms, hemispherical shapes become dominant.

Overall, these tendency is consistent with the results reported by~\citet{Sugiura2018}, although the boundary for elongated or bilobed shapes observed at $v_{\mathrm{imp}} = 150$\ms in the present study corresponds to the dotted line ($v_{\mathrm{imp}}=1.6v_{\mathrm{esc}}$) in~\citet{Sugiura2018}. This shift does not necessarily indicate a substantial change in the physical morphology, because the two classifications differ only in their mass-ratio criterion:
$M_{\mathrm{lr}}/M_{\mathrm{target}}=2.0$ for bilobed and $<2.0$ for elongated shapes. The shift may instead reflect differences in the amount of material escaping or failing to re-accrete within the simulation time, potentially arising from differences in the MPM and SPH schemes, or the time-integration treatment including gravity. We plan to investigate this in future work.

\section{Discussions and Summary}\label{sec:discussion}

In our simulations, a majority of the particles transition into a granular state during the early stages of the impact process as shown in Figure~\ref{fig:irregular_shape}. Consequently, the validity of the yield criterion and the corresponding return mapping scheme serves as the primary factor governing the simulation results, rather than the detailed behavior of the fracture model. Furthermore, in the current system, gravity and its associated frictional forces play a more dominant role than pressure. The present results demonstrate that the elastic and plastic responses of the material, including the pressure evaluation via the Tillotson EOS and the associated frictional treatment, are handled adequately within the MPM framework. The distinguishing characteristics of MPM relative to SPH---namely its robustness against particle irregularity and its ability to suppress tensile instability---become relevant primarily under high-strain-rate or strongly dynamic conditions, such as shock compression and the subsequent tensile (spall) regime, which were not prominently probed in the present simulations~\citep[but see][for the solution in Godunov SPH]{Sugiura2016}.

For the implementation of the DP model, we adopt logarithmic strain (Hencky strain) developed in computer graphics-based MPM for handling large deformations, although the underlying physical principle is shared with conventional methods. This differs from the strain measures commonly employed in traditional SPH frameworks. The resulting agreement with previous studies, despite this difference in strain measure, supports the applicability of MPM to impact simulations within the field of planetary science, at least within the compression-dominated, large-deformation regime examined here.

Future work will involve verifying the applicability of MPM under pressure-dominated conditions, such as hypervelocity impacts exceeding the speed of sound \citep[e.g.,][]{ElMir2019, Yan2026}, or in scenarios where material strength dominates self-gravity~\citep{Benz1999}, as in asteroids with radii of approximately 0.5 km or less \citep[e.g.,][]{Sugiura2019}. Under these conditions, the effects of fracture model are expected to become significant, necessitating further studies. Another important direction is to extend the present model to differentiated targets with core--mantle structure, as considered in SPH simulations \citep[e.g.,][]{Kurosaki2026}. In addition, investigating the shapes of reaccumulated bodies following catastrophic disruption through larger-scale simulations remains an important subject for future research, as previously explored using SPH by \citet{Sugiura2020}.

\begin{acknowledgments}
We thank Yosuke Matsumoto, Hiroshi Kobayashi, Kenji Kurosaki, Shusuke Utsumi, and Riona Yamada for helpful discussions and comments. 
This work has been supported by IAAR Research Support Program in Chiba
University Japan, and MEXT/JSPS KAKENHI Grant Number JP26H02062 (T.~I.) and JP25H00394 (S.~I.).
Numerical computations were carried out on GPU cluster at the Center for Computational Astrophysics, National Astronomical Observatory of Japan.
\end{acknowledgments}


%


\software{
  CUDA~\url{https://developer.nvidia.com/cuda-toolkit},
  Thrust~\url{https://nvidia.github.io/cccl/unstable/thrust/index.html},
  CUB~\url{https://nvidia.github.io/cccl/unstable/cub/index.html},
  PyVista~\citep{sullivan2019pyvista},
  3x3\_SVD\_CUDA~\citep{Gao2018b}
}


\appendix

\section{Correspondence with SPH Governing Equations}\label{sec:CS}

To relate the present MPM/ASFLIP formulation to the standard SPH equations for elastic bodies, we summarize below the correspondence for the time evolution of density, velocity, and energy:

\begin{align}
\rho_p^{n+1} &= \frac{m_p}{\det(\mathit{F}_p^{n+1}) |\Omega_p^{0}|}, \qquad
\mathit{F}_p^{n+1} = \left(\mathit{I} + dt\, \nabla \bm v_p^{n+\frac{1}{2}}\right)\mathit{F}_p^{n},
&&\text{(cf. SPH's } D\rho/Dt \text{)} \label{eq:rho} \\[4pt]
\bm v_i^{n+\frac12} &= \bm v_i^{n-\frac12} + dt\, \frac{\bm f_i^n}{m_i^n},
\qquad
\bm f_i^n = -\sum_p \frac{4}{dx^2}\, |\Omega_p^{0}|\, \mathit{\tau}_p^{n} \, (\bm x_i - \bm x_p^n) w_{ip},
&&\text{(grid, cf. SPH's } D\mathbf{v}/Dt \text{)} \label{eq:v_grid} \\[4pt]
\bm v_p^{n+\frac12} &= (1-\alpha) \sum_i w_{ip}\, \bm v_i^{n+\frac12}
+ \alpha \left(\bm v_p^{n-\frac12} + \sum_i w_{ip}\left(\bm v_i^{n+\frac12} - \bm v_i^{n-\frac12}\right)\right),
&&\text{(particle, reconstruction)} \label{eq:v_particle} \\[4pt]
e_p^{n+1} &= e_p^{n} + \tfrac{dt}{2}\left(\mathit{\tau}_p^{n+1} + \mathit{\tau}_p^{n}\right) : \tfrac{1}{2}\left(\nabla \bm v_p^{n+\frac{1}{2}} + \nabla \bm v_p^{n+\frac{1}{2}, \mathrm T}\right).
&&\text{(cf. SPH's } Du/Dt \text{)} \label{eq:energy}
\end{align}

Here, $m_p$, $\rho_p$, $\mathit{F}_p$, $\nabla \bm v_p$, $\bm x_p$, and $|\Omega_p^{0}|$ are the mass, density, deformation gradient, velocity gradient, position, and reference volume of particle~$p$; $\bm v_i$, $\bm f_i$, and $\bm x_i$ are the grid velocity, force, and position at node $i$, evaluated at leapfrog half-steps; $w_{ip}$ is the particle--grid interpolation weight; $\alpha$ is the ASFLIP blending parameter; $e_p$ and $\mathit{\tau}_p$ are the internal energy density (per reference volume) and Kirchhoff stress, and the corresponding specific internal energy (per unit mass), $E_p = e_p |\Omega_p^{0}| / m_p$, is used. The force $\bm f_i$ follows MLS-MPM~\citep{Hu2018}, which retains the APIC velocity transfer~\citep{Jiang2015}, but reformulates the stress-divergence term via moving least squares, with $\frac{4}{dx^2}$ arising from the quadratic B-spline kernel on a grid of spacing $dx$ (i.e., the MLS-MPM term $\frac{4}{dx^2}(\bm x_i - \bm x_p) w_{ip}$ corresponds to the APIC form $\nabla w_{ip}$).

\section{Stress Model}\label{sec:PS}
The deviatoric Kirchhoff stress tensor is calculated as
\begin{equation}
    \mathit{\tau}^{n}_{\mathrm{dev},p}=2\mu\left(
    \mathit{h}^{\,\mathrm{e},n}_p-\frac{\mathrm{tr}(\mathit{h}^{\,\mathrm{e},n}_p)}{3}\mathit{I}
    \right),
\end{equation}
where $\mathit{h}^{\,\mathrm{e},n}_p=\frac{1}{2}\ln\left(\mathit{b}^{\,\mathrm{e},n}_p\right)$ is the elastic Hencky strain tensor, $\mathit{b}^{\,\mathrm{e},n}_p$ is the elastic left Cauchy--Green tensor, given by $\mathit{b}^{\,\mathrm{e},n}_p=\mathit{F}^{\,\mathrm{e},n}_{p}\mathit{F}^{\,\mathrm{e},n,T}_{p}$, where $\mathit{F}^{\,\mathrm{e},n}_{p}$ denotes the elastic deformation gradient, $\mu$ is the shear modulus, and $\mathit{I}$ is the identity tensor.

The pressure $P_p^{n}$ is given by the sum of the thermodynamic pressure from the Tillotson EOS and the artificial viscosity pressure, further reduced by damage when in tension,
\begin{equation}
\hat{P}_p^n = p_{\mathrm{til}}(E_{p}^{n-1},|\Omega_p^{n}|)+p_{\mathrm{visc}}, \qquad
P_p^{n}=
\begin{cases}
\hat{P}_p^n, & \hat{P}_p^n \ge 0, \\
(1-D_p^n)\,\hat{P}_p^n, & \hat{P}_p^n < 0,
\end{cases}
\end{equation}
where $p_{\mathrm{til}}$ denotes the Tillotson pressure~\citep{Tillotson1962}, $E_p^{n-1}$ is the specific internal energy per unit mass, $|\Omega_p^{n}|$ denotes the current volume of the particle, given by $|\Omega_p^{n}| = J^{n}_{p}|\Omega_p^{0}|$ with $J^{n}_{p} = \det(\mathit{F}^{n}_{p})$, evaluated from the total deformation gradient tensor $\mathit{F}^{n}_{p}$ rather than its elastic part $\mathit{F}^{\,\mathrm{e},n}_{p}$, and $p_{\mathrm{visc}}$ is the artificial viscosity pressure (see Appendix~\ref{sec:visc}). $D_p^n \in [0,1]$ is the damage variable. $E_p^{n-1}$ is obtained from the energy density $e_p^{n-1}$, which is time-integrated by Eq.~\eqref{eq:energy}.
Finally, the Kirchhoff stress tensor is obtained as
\begin{equation}
    \mathit{\tau}^{n}_{p}
    =
    \mathit{\tau}^{n}_{\mathrm{dev},p}
    -
    J^{n}_{p}
    P^{n}_{p}\mathit{I}.
\end{equation}

\section{Artificial Viscosity}\label{sec:visc}
The implementation of artificial viscosity is essential to handle shock waves for impact simulations. Following a previous study employing MPM~\citep{Tonge2016} based on the method proposed in \citet{Wilkins1980}, we add the following artificial viscosity to the pressure term:
\begin{equation}
  p_{\mathrm{visc}} = 
  \begin{cases}
    \rho (A_1C_0 |\tr(\mathit{d})|dx + A_2 \tr(\mathit{d})^2dx^2) & \mathrm{if} \, \tr(\mathit{d}) \le 0\mathrm{,} \\
    0 & \mathrm{if} \, \mathrm{otherwise,}
  \end{cases}
\end{equation}
where $dx$ denotes the cell width of the background grid, $\rho = m/(|\Omega^0|\det(\mathit{F}))$ is the current material density, $C_0$ is the bulk sound speed, and $\mathit{d}$ is the symmetric part of the velocity gradient. We set the parameters $A_1=1.0$ and $A_2=2.0$.

\section{Rock Model and Plasticity Treatment}\label{sec:AP}

Here, we outline the computational algorithm. Let the trial elastic deformation gradient be decomposed by singular value decomposition (SVD) as
\begin{align}
\hat{\mathit{F}}^{\,\mathrm{e}}=\mathit{U}\,\hat{\mathit{\Sigma}}^{\,\mathrm{e}}\,\mathit{V}^{\mathrm T},
\end{align}
where $\hat{\mathit{\Sigma}}^{\,\mathrm{e}}=\mathrm{diag}(\hat{\sigma}_1,\hat{\sigma}_2,\hat{\sigma}_3)$
contains the principal stretches, and $\mathit U$ and $\mathit V$ are orthogonal matrices. 
As a measure of deformation, we use the Hencky strain (logarithmic strain). 
The corresponding trial elastic principal Hencky strain is then defined as
\begin{align}
\hat{\mathit{H}}^{\,\mathrm{e}}=\ln\hat{\mathit{\Sigma}}^{\,\mathrm{e}}=\left(
\ln\hat{\sigma}_1,
\ln\hat{\sigma}_2,
\ln\hat{\sigma}_3
\right)^{\mathrm T}.
\end{align}
The present constitutive model employs the standard radial return mapping on
the smooth portion of the yield surface; detailed derivations are available
in the computational plasticity literature~\citep[e.g.,][]{Klar2016}.
Specifically, the plastic correction via the return mapping in the principal
strain space is performed to obtain the elastically admissible Hencky strain $\mathit{H}^{\,\mathrm{e}}$ as follows:
\begin{align}
\mathit{H}^{\,\mathrm{e}}=\hat{\mathit{H}}^{\,\mathrm{e}}-\delta\gamma\frac{\vb{dev}(\hat{\mathit{H}}^{\,\mathrm{e}})}{\norm{\vb{dev}(\hat{\mathit{H}}^{\,\mathrm{e}})}_{\mathrm{F}}},
\end{align}
where $\vb{dev}(\mathit{A})=\mathit{A}-\frac{1}{3}\tr(\mathit{A})I$ denotes the deviatoric part of a tensor, with $\frac{1}{3}$ being the reciprocal of the spatial dimension, $\delta\gamma$ denotes the plastic multiplier, and $\norm{\cdot}_{\mathrm{F}}$ is the Frobenius norm. 

To determine $\delta\gamma$, the yield strength $Y$ of the rock is required. Following the pressure-dependent strength model of \citet{Lundborg1968}, adapted by \citet{Collins2004, Jutzi2015}, the yield strength $Y$ of the rock in an intermediate state—ranging from intact to granular (with yield strengths $Y_{\mathrm{c}}$ and $Y_{\mathrm{d}}$, respectively)—is determined using the damage variable $D$ as follows:
\begin{align}
Y_{\mathrm{c}} &= \frac{(1-E/E_{\mathrm{melt}})}{2\mu} \left( \sqrt{2}Y_{\mathrm{0}} + \frac{3J P\alpha_{\mathrm{c}}}{1 + \frac{3J^{\,\mathrm{e}} P\alpha_{\mathrm{c}}}{\sqrt{2}(Y_{\mathrm{m}} - Y_{\mathrm{0}})}} \right), \\
Y_{\mathrm{d}} &= \frac{1}{2\mu} 3J P\alpha_{\mathrm{d}}, \\
Y &= (1-D)Y_{\mathrm{c}} + DY_{\mathrm{d}}, \label{eq:DamageLinear}\\
\delta\gamma &= \norm{\vb{dev}(\hat{\mathit H}^{\,\mathrm{e}})}_F - Y.
\end{align}
Here, $\mu$ is the shear modulus, $E_{\mathrm{melt}}$ is the specific melting energy, and $Y_{\mathrm{0}}$ and $Y_{\mathrm{m}}$ denote the shear strengths at $P=0$ and $P\rightarrow\infty$, respectively. The parameters $\mu$ and $E_{\mathrm{melt}}$ are set according to Table I of \citet{Benz1999}, while $Y_{\mathrm{0}}$ and $Y_{\mathrm{m}}$ follow the values specified in~\citet{Jutzi2015}. 
The coefficients $\alpha_{\mathrm{c}}$ and $\alpha_{\mathrm{d}}$ are the Drucker--Prager friction parameters corresponding to friction angles $\phi_{\mathrm{c}}$ and $\phi_{\mathrm{d}}$, respectively, where $\tan\phi_{\mathrm{c}}=1.5$ following~\citet{Jutzi2015}, and $\phi_{\mathrm{d}}=40^\circ$ following~\citet{Sugiura2018}. These are converted from the friction angles via:
\begin{align}
  \alpha_{*}
  =
  \sqrt{\frac{2}{3}}
  \frac{2\sin\phi_{*}}
       {3-\sin\phi_{*}},
\end{align}
for the inner Drucker--Prager approximation. Since strain hardening is not considered in the present study, $\alpha_{\mathrm{*}}$ is treated as a constant.

If $Y\le0$ at the trial value of $JP$, $JP$ is instead solved from $Y=0$ and
used in place of the trial value when constructing the volumetric part of
the corrected Kirchhoff stress tensor, and $\mathit{H}^{\,\mathrm{e}}$
simultaneously reset to $\mathbf{0}$. If $\delta\gamma\le0$, the state is
elastic and $\mathit H^{\,\mathrm{e}}=\hat{\mathit H}^{\,\mathrm{e}}$ is
maintained. Otherwise, $\mathit H^{\,\mathrm{e}}$ is corrected by the radial
return mapping described above. In every case, the corrected principal
stretches are reconstructed from $\mathit H^{\,\mathrm{e}}$, and the
plasticity-corrected elastic and plastic deformation gradients are recovered
as
\begin{align}
\mathit{F}^{\,\mathrm{e}} &= \mathit{U}\exp(\mathit{H}^{\,\mathrm{e}})\mathit{V}^T, \\
\mathit{F}^{\,\mathrm{p}} &= \mathit{V}\exp(-\mathit{H}^{\,\mathrm{e}})\exp\small(\hat{\mathit{H}}^{\,\mathrm{e}})\mathit{V}^T\hat{\mathit{F}}^{\,\mathrm{p}},
\end{align}
where $\mathit U$ and $\mathit V$ are identical to those obtained from the
trial elastic deformation gradient owing to the coaxial return mapping, and
$\hat{\mathit{F}}^{\,\mathrm{p}}$, the plastic deformation gradient entering
this correction, remains unchanged from the previous time step since only
$\hat{\mathit{F}}^{\,\mathrm{e}}$ is updated through the G2P step. The total
deformation gradient $\mathit{F}=\mathit{F}^{\,\mathrm{e}}\mathit{F}^{\,\mathrm{p}}$
thus remains unchanged by the plastic correction, which only redistributes
the deformation between its elastic and plastic parts.


\bibliographystyle{aasjournalv7}

\begin{thebibliography}{}
\expandafter\ifx\csname natexlab\endcsname\relax\def\natexlab#1{#1}\fi
\providecommand{\url}[1]{\href{#1}{#1}}
\providecommand{\dodoi}[1]{doi:~\href{http://doi.org/#1}{\nolinkurl{#1}}}
\providecommand{\doeprint}[1]{\href{http://ascl.net/#1}{\nolinkurl{http://ascl.net/#1}}}
\providecommand{\doarXiv}[1]{\href{https://arxiv.org/abs/#1}{\nolinkurl{https://arxiv.org/abs/#1}}}
\bibitem[{W. Benz \& E. Asphaug(1995)Benz \& Asphaug}]{Benz1995}
Benz, W., \& Asphaug, E. 1995, \bibinfo{title}{Simulations of brittle solids
  using smooth particle hydrodynamics,} Computer Physics Communications, 87,
  253, \dodoi{https://doi.org/10.1016/0010-4655(94)00176-3}

\bibitem[{W. Benz \& E. Asphaug(1999)Benz \& Asphaug}]{Benz1999}
Benz, W., \& Asphaug, E. 1999, \bibinfo{title}{Catastrophic Disruptions
  Revisited,} Icarus, 142, 5, \dodoi{https://doi.org/10.1006/icar.1999.6204}

\bibitem[{G.~S. Collins {et~al.}(2004)Collins, Melosh, \& Ivanov}]{Collins2004}
Collins, G.~S., Melosh, H.~J., \& Ivanov, B.~A. 2004, \bibinfo{title}{Modeling
  damage and deformation in impact simulations,} Meteoritics \& Planetary
  Science, 39, 217, \dodoi{https://doi.org/10.1111/j.1945-5100.2004.tb00337.x}

\bibitem[{D.~C. Drucker \& W. Prager(1952)Drucker \& Prager}]{Drucker1952}
Drucker, D.~C., \& Prager, W. 1952, \bibinfo{title}{Soil mechanics and plastic
  analysis or limit design,} Quarterly of Applied Mathematics, 10, 157.
\newblock \url{https://api.semanticscholar.org/CorpusID:39707105}

\bibitem[{C. {El Mir} {et~al.}(2019){El Mir}, Ramesh, \&
  Richardson}]{ElMir2019}
{El Mir}, C., Ramesh, K., \& Richardson, D.~C. 2019, \bibinfo{title}{A new
  hybrid framework for simulating hypervelocity asteroid impacts and
  gravitational reaccumulation,} Icarus, 321, 1013,
  \dodoi{https://doi.org/10.1016/j.icarus.2018.12.032}

\bibitem[{Y.~R. Fei {et~al.}(2021)Fei, Guo, Wu, Huang, \& Gao}]{Fei2021}
Fei, Y.~R., Guo, Q., Wu, R., Huang, L., \& Gao, M. 2021,
  \bibinfo{title}{Revisiting integration in the material point method: a scheme
  for easier separation and less dissipation,} ACM Trans. Graph., 40,
  \dodoi{10.1145/3450626.3459678}

\bibitem[{C. Fu {et~al.}(2017)Fu, Guo, Gast, Jiang, \& Teran}]{Fu2017}
Fu, C., Guo, Q., Gast, T., Jiang, C., \& Teran, J. 2017, \bibinfo{title}{A
  polynomial particle-in-cell method,} ACM Trans. Graph., 36,
  \dodoi{10.1145/3130800.3130878}

\bibitem[{M. Gao {et~al.}(2018)Gao, Wang, Wu, Pradhana, Sifakis, Yuksel, \&
  Jiang}]{Gao2018b}
Gao, M., Wang, X., Wu, K., {et~al.} 2018, \bibinfo{title}{GPU optimization of
  material point methods,} ACM Trans. Graph., 37,
  \dodoi{10.1145/3272127.3275044}

\bibitem[{D. Grady \& M. Kipp(1980)Grady \& Kipp}]{Grady1980}
Grady, D., \& Kipp, M. 1980, \bibinfo{title}{Continuum modelling of explosive
  fracture in oil shale,} International Journal of Rock Mechanics and Mining
  Sciences \& Geomechanics Abstracts, 17, 147,
  \dodoi{https://doi.org/10.1016/0148-9062(80)91361-3}

\bibitem[{Y. Hu {et~al.}(2018)Hu, Fang, Ge, Qu, Zhu, Pradhana, \&
  Jiang}]{Hu2018}
Hu, Y., Fang, Y., Ge, Z., {et~al.} 2018, \bibinfo{title}{A moving least squares
  material point method with displacement discontinuity and two-way rigid body
  coupling,} ACM Trans. Graph., 37, \dodoi{10.1145/3197517.3201293}

\bibitem[{J.~P. {Huchra} \& M.~J. {Geller}(1982){Huchra} \&
  {Geller}}]{Huchra1982}
{Huchra}, J.~P., \& {Geller}, M.~J. 1982, \bibinfo{title}{{Groups of Galaxies.
  I. Nearby groups},} \apj, 257, 423, \dodoi{10.1086/160000}

\bibitem[{C. Jiang {et~al.}(2015)Jiang, Schroeder, Selle, Teran, \&
  Stomakhin}]{Jiang2015}
Jiang, C., Schroeder, C., Selle, A., Teran, J., \& Stomakhin, A. 2015,
  \bibinfo{title}{The affine particle-in-cell method,} ACM Trans. Graph., 34,
  \dodoi{10.1145/2766996}

\bibitem[{M. Jutzi(2015)Jutzi}]{Jutzi2015}
Jutzi, M. 2015, \bibinfo{title}{SPH calculations of asteroid disruptions: The
  role of pressure dependent failure models,} Planetary and Space Science, 107,
  3, \dodoi{https://doi.org/10.1016/j.pss.2014.09.012}

\bibitem[{M. Jutzi \& E. Asphaug(2015)Jutzi \& Asphaug}]{Jutzi2015b}
Jutzi, M., \& Asphaug, E. 2015, \bibinfo{title}{The shape and structure of
  cometary nuclei as a result of low-velocity accretion,} Science, 348, 1355,
  \dodoi{10.1126/science.aaa4747}

\bibitem[{ {Jutzi, M.} \&  {Benz, W.}(2017){Jutzi, M.} \& {Benz,
  W.}}]{Jutzi2017}
{Jutzi, M.}, \& {Benz, W.} 2017, \bibinfo{title}{Formation of bi-lobed shapes
  by sub-catastrophic collisions - A late origin of comet 67P�fs structure,}
  A\&A, 597, A62, \dodoi{10.1051/0004-6361/201628964}

\bibitem[{G. Kl\'{a}r {et~al.}(2016)Kl\'{a}r, Gast, Pradhana, Fu, Schroeder,
  Jiang, \& Teran}]{Klar2016}
Kl\'{a}r, G., Gast, T., Pradhana, A., {et~al.} 2016,
  \bibinfo{title}{Drucker-prager elastoplasticity for sand animation,} ACM
  Trans. Graph., 35, \dodoi{10.1145/2897824.2925906}

\bibitem[{K. Kurosaki \& M. Arakawa(2026)Kurosaki \& Arakawa}]{Kurosaki2026}
Kurosaki, K., \& Arakawa, M. 2026, \bibinfo{title}{Reaccumulation process after
  a catastrophic disruption event on a differentiated asteroid,} Icarus, 457,
  117167, \dodoi{https://doi.org/10.1016/j.icarus.2026.117167}

\bibitem[{N. Lundborg(1968)Lundborg}]{Lundborg1968}
Lundborg, N. 1968, \bibinfo{title}{Strength of rock-like materials,}
  International Journal of Rock Mechanics and Mining Sciences \& Geomechanics
  Abstracts, 5, 427, \dodoi{https://doi.org/10.1016/0148-9062(68)90046-6}

\bibitem[{S. Ma {et~al.}(2009)Ma, Zhang, \& Qiu}]{Ma2009}
Ma, S., Zhang, X., \& Qiu, X. 2009, \bibinfo{title}{Comparison study of MPM and
  SPH in modeling hypervelocity impact problems,} International Journal of
  Impact Engineering, 36, 272,
  \dodoi{https://doi.org/10.1016/j.ijimpeng.2008.07.001}

\bibitem[{C. Mast {et~al.}(2015)Mast, Arduino, Mackenzie-Helnwein, \&
  Miller}]{Mast2014}
Mast, C., Arduino, P., Mackenzie-Helnwein, P., \& Miller, R. 2015,
  \bibinfo{title}{Simulating granular column collapse using the material point
  method,} Acta Geotechnica, 10, 101, \dodoi{10.1007/s11440-014-0309-0}

\bibitem[{C.~M. Mast(2013)Mast}]{Mast2013}
Mast, C.~M. 2013, \bibinfo{title}{Modeling Landslide-Induced Flow Interactions
  with Structures using the Material Point Method,} PhD thesis, University of
  Washington.
\newblock \url{http://hdl.handle.net/1773/23580}

\bibitem[{ {Michel, P.} \&  {Richardson, D. C.}(2013){Michel, P.} \&
  {Richardson, D. C.}}]{Michel2013}
{Michel, P.}, \& {Richardson, D. C.} 2013, \bibinfo{title}{Collision and
  gravitational reaccumulation: Possible formation mechanism of the asteroid
  Itokawa,} A\&A, 554, L1, \dodoi{10.1051/0004-6361/201321657}

\bibitem[{D.~P. O'Brien \& R. Greenberg(2005)O'Brien \&
  Greenberg}]{O'Brien2005}
O'Brien, D.~P., \& Greenberg, R. 2005, \bibinfo{title}{The collisional and
  dynamical evolution of the main-belt and NEA size distributions,} Icarus,
  178, 179, \dodoi{https://doi.org/10.1016/j.icarus.2005.04.001}

\bibitem[{S.~R. Schwartz {et~al.}(2018)Schwartz, Michel, Jutzi, Marchi, Zhang,
  \& Richardson}]{Schwartz2018}
Schwartz, S.~R., Michel, P., Jutzi, M., {et~al.} 2018,
  \bibinfo{title}{Catastrophic disruptions as the origin of bilobate comets,}
  Nature Astronomy, 2, 379, \dodoi{10.1038/s41550-018-0395-2}

\bibitem[{M. Steffen {et~al.}(2008)Steffen, Kirby, \& Berzins}]{Michael2008}
Steffen, M., Kirby, R.~M., \& Berzins, M. 2008, \bibinfo{title}{Analysis and
  reduction of quadrature errors in the material point method (MPM),}
  International Journal for Numerical Methods in Engineering, 76, 922,
  \dodoi{https://doi.org/10.1002/nme.2360}

\bibitem[{K. Sugiura \& S. Inutsuka(2016)Sugiura \& Inutsuka}]{Sugiura2016}
Sugiura, K., \& Inutsuka, S. 2016, \bibinfo{title}{An extension of Godunov SPH:
  Application to negative pressure media,} Journal of Computational Physics,
  308, 171, \dodoi{https://doi.org/10.1016/j.jcp.2015.12.030}

\bibitem[{K. Sugiura {et~al.}(2018)Sugiura, Kobayashi, \&
  Inutsuka}]{Sugiura2018}
Sugiura, K., Kobayashi, H., \& Inutsuka, S. 2018, \bibinfo{title}{Toward
  understanding the origin of asteroid geometries - Variety in shapes produced
  by equal-mass impacts,} A\&A, 620, A167, \dodoi{10.1051/0004-6361/201833227}

\bibitem[{K. Sugiura {et~al.}(2019)Sugiura, Kobayashi, \&
  Inutsuka}]{Sugiura2019}
Sugiura, K., Kobayashi, H., \& Inutsuka, S. 2019, \bibinfo{title}{Collisional
  elongation: Possible origin of extremely elongated shape of 1I/�eOumuamua,}
  Icarus, 328, 14, \dodoi{https://doi.org/10.1016/j.icarus.2019.03.014}

\bibitem[{K. Sugiura {et~al.}(2020)Sugiura, Kobayashi, \&
  Inutsuka}]{Sugiura2020}
Sugiura, K., Kobayashi, H., \& Inutsuka, S. 2020,
  \bibinfo{title}{High-resolution simulations of catastrophic disruptions:
  Resultant shape distributions,} Planetary and Space Science, 181, 104807,
  \dodoi{https://doi.org/10.1016/j.pss.2019.104807}

\bibitem[{C.~B. Sullivan \& A.~A. Kaszynski(2019)Sullivan \&
  Kaszynski}]{sullivan2019pyvista}
Sullivan, C.~B., \& Kaszynski, A.~A. 2019, \bibinfo{title}{PyVista: 3D plotting
  and mesh analysis through a streamlined interface for the Visualization
  Toolkit (VTK),} Journal of Open Source Software, 4, 1450,
  \dodoi{10.21105/joss.01450}

\bibitem[{D. Sulsky {et~al.}(1995)Sulsky, Zhou, \& Schreyer}]{Sulsky1995}
Sulsky, D., Zhou, S.-J., \& Schreyer, H.~L. 1995, \bibinfo{title}{Application
  of a particle-in-cell method to solid mechanics,} Computer Physics
  Communications, 87, 236, \dodoi{https://doi.org/10.1016/0010-4655(94)00170-7}

\bibitem[{J.~H. Tillotson(1962)Tillotson}]{Tillotson1962}
Tillotson, J.~H. 1962, \bibinfo{title}{Metallic Equations of State for
  Hypervelocity Impact. GA-3216,} General Atomic Division of General Dynamics.
\newblock \url{https://cir.nii.ac.jp/crid/1571980074126894592}

\bibitem[{A.~L. Tonge \& K. Ramesh(2016)Tonge \& Ramesh}]{Tonge2016}
Tonge, A.~L., \& Ramesh, K. 2016, \bibinfo{title}{Multi-scale defect
  interactions in high-rate brittle material failure. Part I: Model formulation
  and application to ALON,} Journal of the Mechanics and Physics of Solids, 86,
  117, \dodoi{https://doi.org/10.1016/j.jmps.2015.10.007}

\bibitem[{M.~L. Wilkins(1980)Wilkins}]{Wilkins1980}
Wilkins, M.~L. 1980, \bibinfo{title}{Use of artificial viscosity in
  multidimensional fluid dynamic calculations,} Journal of Computational
  Physics, 36, 281, \dodoi{https://doi.org/10.1016/0021-9991(80)90161-8}

\bibitem[{X. Yan {et~al.}(2026)Yan, Michel, Ni, Jiao, \& Li}]{Yan2026}
Yan, X., Michel, P., Ni, R., Jiao, Y., \& Li, J. 2026, \bibinfo{title}{The
  Material Point Method (MPM) for simulating hypervelocity impact on
  asteroids,} Icarus, 455, 117080, \dodoi{10.1016/j.icarus.2026.117080}

\bibitem[{Y. Zhu \& R. Bridson(2005)Zhu \& Bridson}]{Zhu2005}
Zhu, Y., \& Bridson, R. 2005, \bibinfo{title}{Animating sand as a fluid,} ACM
  Trans. Graph., 24, 965^^e2^^80^^93972, \dodoi{10.1145/1073204.1073298}

\end{thebibliography}



\end{document}